\documentclass{article}
\usepackage{spconf,amssymb,amsmath,graphicx,times,url}
\usepackage{color}
\usepackage{subfig}

\usepackage{bm}
\usepackage{amsthm}
\usepackage{amsfonts,amscd,mathrsfs}
\usepackage{dsfont}
\usepackage[mathscr]{euscript}
\usepackage{latexsym}

\usepackage{fakufakumath}

\usepackage{layouts}

\usepackage{xcolor}
\definecolor{codegreen}{rgb}{0,0.6,0}
\definecolor{codegray}{rgb}{0.5,0.5,0.5}
\definecolor{codepurple}{rgb}{0.58,0,0.82}
\definecolor{backcolour}{rgb}{0.95,0.95,0.92}

\usepackage{listings}
\lstdefinestyle{mystyle}{
    backgroundcolor=\color{backcolour},
    commentstyle=\color{codegreen},
    keywordstyle=\color{magenta},
    numberstyle=\tiny\color{codegray},
    stringstyle=\color{codepurple},
    basicstyle=\scriptsize,
    breakatwhitespace=false,
    breaklines=true,
    captionpos=b,
    keepspaces=true,
    numbers=left,
    numbersep=5pt,
    showspaces=false,
    showstringspaces=false,
    showtabs=false,
    tabsize=2
}

\title{PYROOMACOUSTICS: A PYTHON PACKAGE FOR AUDIO ROOM SIMULATION AND ARRAY PROCESSING ALGORITHMS}

\graphicspath{{}{figures/}}

\name{Robin Scheibler,$^{1}$ %\sthanks{Thanks to ABC agency for funding.}
      Eric Bezzam,$^{1}$ %\sthanks{Thanks to XYZ agency for funding.}
      Ivan Dokmani\'{c},$^{2}$%\sthanks{Also many thanks.}
    }
\address{%
  $^1$\'{E}cole Polytechnique F\'{e}d\'{e}rale de Lausanne (EPFL), Switzerland \\%
  $^2$University of Illinois Urbana-Champaign, USA \\%
  robin.scheibler@epfl.ch, eric.bezzam@epfl.ch, dokmanic@illinois.edu%
}

\begin{document}

\ninept
\maketitle

\begin{sloppy}

\begin{abstract}
  We present \textit{pyroomacoustics}, a software package aimed at the rapid
  development and testing of audio array processing algorithms.  The content of
  the package can be divided into three main components: an intuitive Python
  object-oriented interface to quickly construct different simulation scenarios
  involving multiple sound sources and microphones in 2D and 3D rooms; a fast
  C~implementation of the image source model for general polyhedral rooms to
  efficiently generate room impulse responses and simulate the propagation
  between sources and receivers; and finally, reference implementations of
  popular algorithms for beamforming, direction finding, and adaptive
  filtering.  Together, they form a package with the potential to speed up the
  time to market of new algorithms by significantly reducing the implementation
  overhead in the performance evaluation step.
\end{abstract}


%\printinunitsof{cm}\prntlen{\linewidth}

\section{Introduction}

As in all engineering disciplines, objective evaluation of new (array) audio
processing algorithms is essential to the assessment of their value.  The gold
standard for such evaluation is to design and carry out an experiment in a
controlled environment with a real microphone array and careful calibration of
the locations of all sound sources. The time and effort needed to setup these
experiments naturally limit the number of replications of the experiments and
the number of scenarios that can be explored. In the exploratory phase of
research, numerical simulation is an attractive alternative. It allows one to
quickly test and iterate a large number of ideas. In addition it makes it possible to
finely tune parameters for the algorithm before going to experiments in the
physical world.

There are typically three components in a simulation system. First, a
programming language to describe the scenarios to simulate. Second, a computer
program implementing a model that simulates the relevant physical effects, in
our case the propagation of sound in air.  Finally, we need computer programs
implementing the algorithms under investigation.

While low level languages like C and FORTRAN are the most efficient when it
comes to speed, they come at a significant cost in terms of implementation
complexity.  High level scripting languages have long been a popular
alternative for describing simulations. 
In particular, MATLAB has been
historically the industry, and often academic, standard for signal processing
numerical experiments. Its high level syntax based on linear algebraic
operations is indeed particularly well suited for algorithms in this field.  It
comes however with significant drawbacks: high cost, closed source, and a clunky
syntax for anything other than linear algebra.  
In recent years Python has come to prominence as an attractive alternative to
MATLAB for high level scientific computing~\cite{Oliphant:2007dm}. Its focus on
code readability and extensibility makes it a candidate of choice for reproducible
scientific code~\cite{lutz2011programming}. The \textit{numpy} and
\textit{scipy} modules extend Python with the same powerful linear algebra
primitives that MATLAB enjoys. An aspect of special interest for DSP engineers
is the massive adoption of Python in the machine learning community as
exemplified by the \textit{scikit-learn} \cite{Pedregosa:2011tv} or speech
recognition packages \cite{zhang2017speech}.  Finally, Python is a community
supported free and open source project that includes robust tools for
distribution (\url{http://pypi.python.org}), documentation
(\url{http://readthedocs.io}), and continuous integration
(\url{http://travis-ci.org}).
% MATLAB cost makes it unavailable with institutions with smaller financial means
% Licensing makes it impractical to run on clusters

%We chose to develop the project in the Python language. Python
%is open source and available for free. It is a modern object oriented language
%that values readability first. It is suitable for scientific computations and
%includes interfaces to standard high performance linear algebra libraries such
%a the linear algebra package (LAPACK), the basic linear algebra subroutine
%(BLAS), or the Intel math kernel library (MKL) \cite{Oliphant:2007dm}.  The
%popularity of Python being not limited to the science and engineering community
%means that a variety of package and extensions are available for tasks from
%database access to web server, making it easy to integrate scientific code in
%applications. For example creating an online demonstration website for a
%research project.

For a simulation to yield useful information and practical insight, it is vital
that it models accurately enough real conditions.  In room acoustics,
simulation based on the image source model (ISM) has been used extensively for this
purpose and has well-known strengths and weaknesses \cite{Allen:1979cn}.  This
model replaces reflections on walls by virtual sources playing the same sound
as the original source and builds a room impulse response (RIR) from the
corresponding delays and attenuations.  Its strength is its simplicity.  The
model is accurate only as long as the wavelength of the sound is small relative
to the size of the reflectors, which it assumes to be uniformly
absorbing across frequencies. Nevertheless, these assumptions are not too far from
reality in many environments of interests such as offices. The original model
can be extended to convex and non-convex polyhedral rooms in two and three
dimensions \cite{Borish:1984uu}.  Our wishlist for an RIR generator is:
affordable, open source, and flexible.
A number of generators are available, and most if not all are shared online
free of charge, e.g.~the popular generator from Emanu\"{e}l Habets
\cite{Habets:2006xy}. Unfortunately, none allow room shapes other than
rectangular. Furthermore, most rely on MATLAB.  Faced by the limitations of
available RIR generators, we decided to develop our own. 

We provide \textit{pyroomacoustics}, a comprehensive Python package for audio
algorithms simulation. The package includes both a fast RIR generator
based on the ISM and a number of reference implementations of
popular algorithms for beamforming, direction of arrival (DOA) finding, and
adaptive filtering. A short time Fourier transform (STFT) engine allows for
efficient frequency domain processing. The object oriented features of Python
are used to provide a LEGO-like interface to these different blocks.  This
paper gives an overview of the usage and content of \textit{pyroomacoustics}.
The software itself is available under a permissive open source license through
the standard Python package manager\footnote{\texttt{pip install
pyroomacoustics}} or on
github\footnote{\url{https://github.com/LCAV/pyroomacoustics}}.
\begin{figure}
  \includegraphics[width=\linewidth]{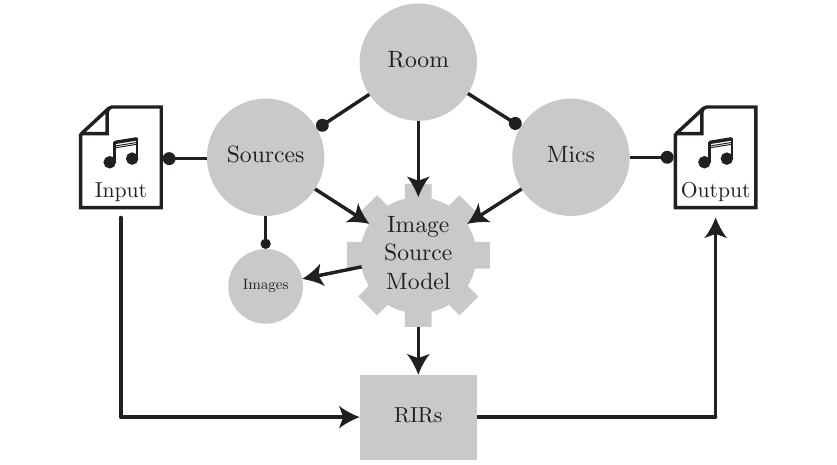}
  \caption{Structure of the sound propagation simulator. The lines terminated
    by a bullet indicate attribute relationship. Arrows indicates parameters to
  functions.}
  \flabel{sim_struct}
\end{figure}

\section{Pyroomacoustics Core}

The \textit{pyroomacoustics} package exploits the object oriented features of
Python to create a clean and intuitive application programming interface (API)
for room acoustics simulation. The three main classes are
\texttt{Room}, \texttt{SoundSource}, and \texttt{MicrophoneArray}.  On a high
level, a simulation scenario is created by first defining a room to which a few
sound sources and a microphone array are attached.  The actual audio is
attached to the source as a raw audio sample.  The ISM is then
used to find all image sources up to a maximum specified order and RIRs are
generated from their positions. The RIR generator is described in more details
in \sref{pyrm_rir_gen}.  The microphone signals are then created by convolving
the audio samples associated to sources with the appropriate RIR. Since the
simulation is done on discrete-time signals, a sampling frequency is specified for
the room and the sources it contains. Microphones can optionally operate at a
different sampling frequency; a rate conversion is done in this case. 
A simple code example and its output are shown in \ffref{pyrm_room_rir}.

\subsection{The Room Class}

A \texttt{Room} object has as attributes a collection of \texttt{Wall} objects,
a microphone array, and a list of sound sources. The room can be two
dimensional (2D), in which case the walls are simply line segments. A
factory method \texttt{from\_corners} can be used to create the room from a
polygon. In three dimensions (3D), the walls are two dimensional polygons, namely a
collection of points lying on a common plane. Creating rooms in 3D is more tedious
and for convenience a method \texttt{extrude} is provided to lift a 2D room
into 3D space by adding vertical walls and a parallel ``ceiling'' (see
\ffref{pyrm_rir}). The \texttt{Room} is sub-classed by \texttt{ShoeBox} which
creates a rectangular (2D) or parallelepipedic (3D) room. As will be detailed
in \sref{pyrm_rir_gen}, such rooms benefit from an efficient algorithm for the ISM.

\subsection{The SoundSource Class}

A \texttt{SoundSource} object has as attributes the locations of the source
itself and also all of its image sources. This list is usually
generated by the \texttt{Room} object containing the source.  The reason for
this structure is to anticipate scenarios where the room is defined by the
locations of the image sources, for example in room inference problems
\cite{Dokmanic:2013dz}.  The source object also contains the methods to
build an RIR from the image sources locations. The image sources are
conveniently available through the overloaded bracket operator. This comes
handy to select only a subset of image sources, such as when building acoustic rake
receivers \cite{Dokmanic:2015dr}.

\subsection{The MicrophoneArray and Beamformer Classes}
\seclabel{pck_micbf}

The \texttt{MicrophoneArray} class is essentially an array of microphone
locations together with a sampling frequency. It has in addition a
\texttt{record} method that wraps potential rate conversions when the simulation
and microphones are at different rates. 

The \texttt{Beamformer} class inherits from \texttt{MicrophoneArray} and can be
used instead. In that case, beamforming weights (in the frequency domain) or
filters (in the time domain) can be computed according to several methods (see
\sref{algo_bf}).  \ffref{ds_example} shows an example of a
delay-and-sum (DS) beamformer in a rectangular room. 

In addition, the \texttt{Beamformer} class packs an STFT engine for efficient
frame processing in the frequency domain (see \ffref{stft}). The engine allows for variable size
zero-padding, overlap, and the use of different analysis and synthesis windows.
Alternatively, direct filtering in the time domain is also possible.
Specialized methods can convert weights to a corresponding filter and
vice-versa. In case of mismatch in size, a least squares fit of the beamforming
weights to a smaller filter is done.

\begin{figure}
\begin{center}
  \includegraphics[width=0.49\textwidth]{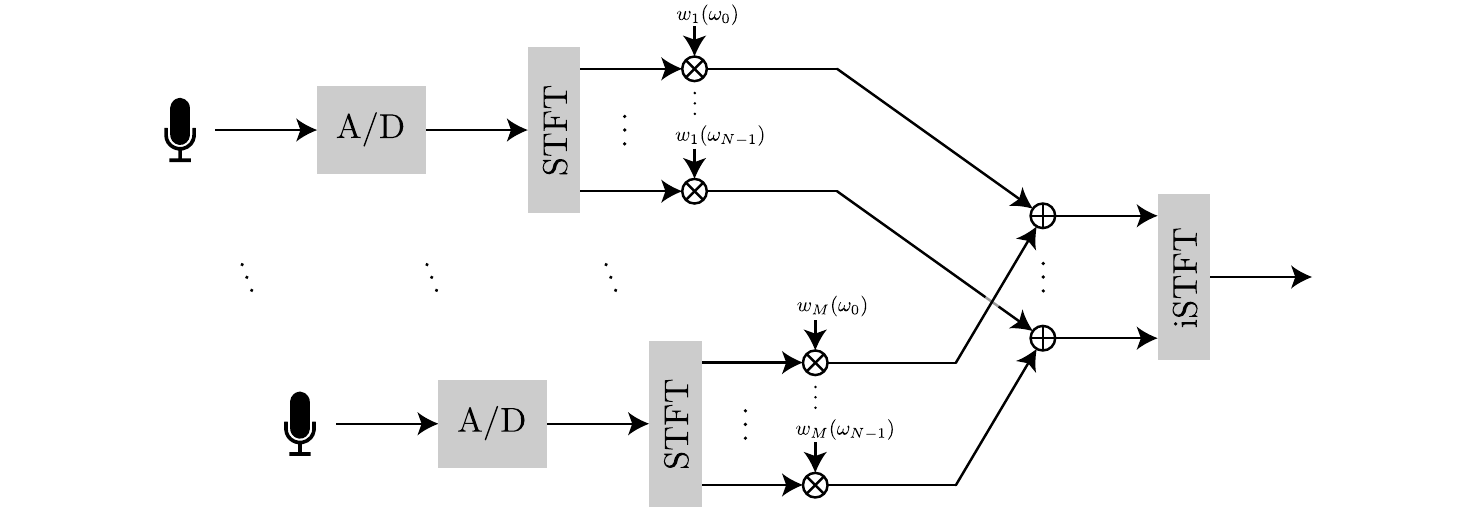}
\end{center}
\caption{Block diagram of STFT domain beamforming.}
\flabel{stft}
\end{figure}

\begin{figure}
\begin{center}
  \includegraphics[width=\linewidth]{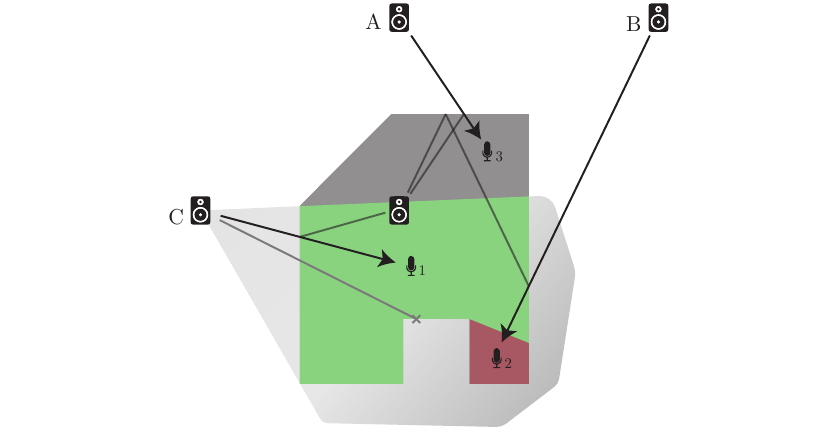}
\end{center}
\caption{Image source model for arbitrary polyhedral rooms. Sources A and B are visible from all microphones. 
  Source C is only visible in the shaded/green area due to the obtuse angle.
  Microphone 2 is hidden due to obstruction by a re-entrant wall.}
\flabel{ism}
\end{figure}

\section{Room Impulse Response Generator}
\seclabel{pyrm_rir_gen}

\begin{figure*}
  \begin{minipage}{0.45\textwidth}
    \begin{center}
      \subfloat[The room]{\flabel{pyrm_room} \includegraphics[width=1.\textwidth]{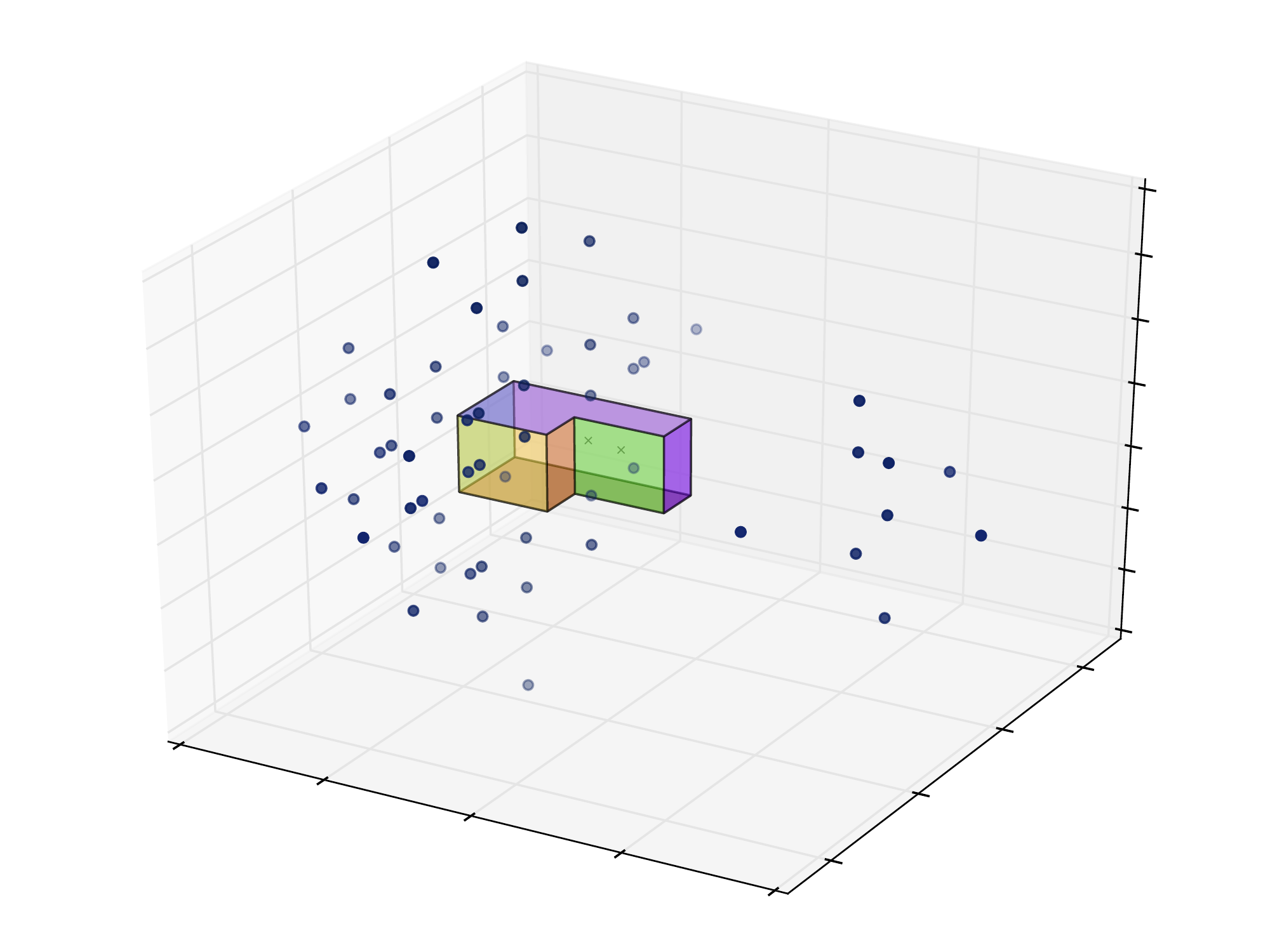}} \\
      \subfloat[The RIRs]{\flabel{pyrm_rir} \includegraphics[width=1.\textwidth]{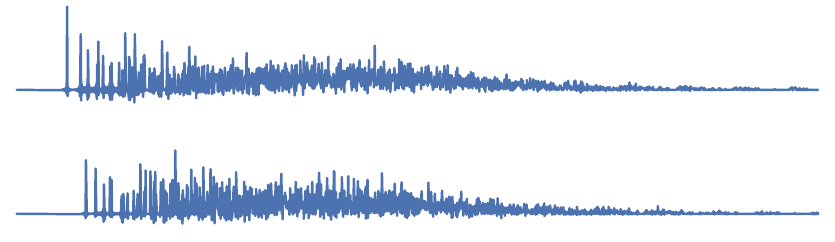}}
    \end{center}
  \end{minipage}
  \hfill
  \subfloat[Python code to generate these figures]{
    \begin{minipage}{0.500\textwidth}
      \flabel{img_src_code}
\begin{lstlisting}[style=mystyle, language=Python]
import numpy as np
import matplotlib.pyplot as plt
import pyroomacoustics as pra

# Create a 2D room from the corners of a polygon
floor = np.array([[0, 0, 6, 6,   3,   3],   # x-coordinates
                  [0, 3, 3, 1.5, 1.5, 0]])  # y-coordinates
room = pra.Room.from_corners(floor, fs=16000, 
                             max_order=12, absorption=0.1)

# Lift the room in 3D space
room.extrude(2.4)

# Add a sound source
room.add_source([1.5, 1.2, 1.6])

# Place two microphones in the room
R = np.array([[3., 4.2], [2.25, 2.1], [1.4, 1.4]])
bf = pra.MicrophoneArray(R, room.fs)
room.add_microphone_array(bf)

# Run image source model and show room and 2nd order images
room.image_source_model()
room.plot(img_order=3, aspect='equal')

fig = plt.figure()
room.plot_rir()
plt.show()
\end{lstlisting}
    \end{minipage}
  }
  \caption{(a) An example of a non-convex room containing one source and two
    microphones with up to 3rd order image sources drawn. (b) The two RIRs between the
    source and the microphones produced by pyroomacoustics.
    (c) The Python code used to generate the first two figures.
  }
  \flabel{pyrm_room_rir}
\end{figure*}

The RIR generator is based on the ISM and considers two cases: shoe box, i.e.
rectangular, and arbitrary polyhedral rooms. For shoe box rooms, the original
algorithm from Allen and Berkley \cite{Allen:1979cn} is used. In this case,
symmetries limit the number of image sources to grow quadratically and
cubically in 2D and 3D, respectively, in the order of reflections. In addition,
image sources are always visible from anywhere in the room. The situation for
arbitrary polyhedral rooms is not that simple. The number of image sources grows
exponentially in the order of reflections and the visibility of
sources must be checked. When obtuse angles occur between walls, the
reflections from these walls will not be visible in the whole room. In
addition, if the room is not convex, i.e. re-entrant walls occur, they
might obstruct the path between image sources and microphones. Both situations
are illustrated in \ffref{ism}. The algorithm is explained in detail in the original
paper by Borish \cite{Borish:1984uu}. In practice, we found its pure Python
implementation to be too slow to be practical and hence moved to compiled C code
for that part of the package.

Once the location of image sources and their visibility from each microphone is
determined, they can be used to construct the RIRs themselves.  For a
microphone placed at $\vr$, a real source $\vs_0$, and a set of its visible
image sources $\mathcal{V}_{\vr}(\vs_0)$, the impulse response between $\vr$
and $\vs_0$, sampled at $F_s$, is given by
\begin{equation}
  \label{eq:rr:rir_lowpass}
  a_{\vr}(\vs_0, n) = 
  \sum_{\vs\in \mathcal{V}_{\vr}(\vs_0)} \frac{(1-\alpha)^{\text{gen}(\vs)}}{4\pi\|\vr-\vs\|}\,\delta_{\text{LP}}\left(n - F_s\frac{\|\vr-\vs\|}{c}\right),
\end{equation}
where $\text{gen}(\vs)$ gives the reflection order of source $\vs$, $\alpha \in
[0,1]$ is the absorption factor of the walls, $c$ is the speed of sound, and
$\delta_\text{LP}$ is the windowed sinc function
\begin{equation}
  \delta_\text{LP}(t) = 
  \begin{cases}
  \frac{1}{2} \left( 1 + \cos \left(\frac{2\pi{t}}{T_w}\right) \right) \sinc(t)& \text{if $-\frac{T_w}{2} \leq t \leq \frac{T_w}{2}$,} \\
    0 & \text{otherwise.}
  \end{cases}
\end{equation}
The parameter $T_w$ controls the width of the window and thus the degree of
approximation to a full sinc. Note that for simplicity we assumed the
absorption factor to be identical for all walls. Nevertheless, the package
allows to specify a different absorption factor for each wall. Two RIRs produced
this way can be seen in \ffref{pyrm_rir}.

\begin{figure*}
    \subfloat[DS Beamformer]{ \flabel{sub_ds_example} \begin{minipage}{0.45\linewidth} \includegraphics[width=\linewidth]{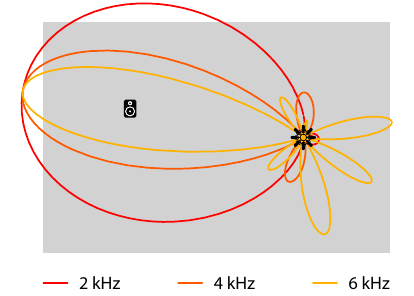} \end{minipage} }
      \hfill
    \subfloat[Python code generating figure (a)]{ \flabel{ds_src_code} 
      \begin{minipage}{0.45\linewidth}
\begin{lstlisting}[style=mystyle, language=Python]
import numpy as np
import matplotlib.pyplot as plt
import pyroomacoustics as pra

# Create a 4 by 6 metres shoe box room
room = pra.ShoeBox([4,6], fs=16000)

# Add a source somewhere in the room
room.add_source([2.5, 4.5])

# circular array with 4 microphones and radius 4 cm
R = pra.circular_2D_array([2, 1.5], 8, 0, 0.04)
bf = pra.Beamformer(R, room.fs)
room.add_microphone_array(bf)

# Now compute the delay and sum weights
room.micArray.rake_delay_and_sum_weights(
                                 room.sources[0][:1])

# plot the room and resulting beamformer
room.plot(freq=[2000, 4000, 6000], img_order=0)
plt.show()
\end{lstlisting}
      \end{minipage}%
    }
  \caption{(a) Beampatterns for a circular delay-and-sum beamformer at 2, 4, and 8 kHz. (b) The code that produced the figure.}
  \flabel{ds_example}
\end{figure*}

\section{Reference Implementations}

When evaluating the performance of new algorithms, a large amount of time is
spent re-implementing competing methods to run comparisons and benchmarks.
While these algorithms are well-known, the devil is always in the details and
their correct practical implementation can be very time consuming. The
availability of robust, tested reference implementations for popular algorithms
has the potential to speed up considerably the time-to-market of new research
projects.  We provide implementations of several algorithms for beamforming,
direction of arrival (DOA) finding, adaptive filtering, and source separation.

\subsection{Beamforming and Source Separation} 
\seclabel{algo_bf}

As described in \sref{pck_micbf}, both frequency and time domain beamformers
can be computed by calling methods from the \texttt{Beamformer} class.  The
classic beamforming algorithms are included as special cases of the acoustic
rake receivers of \cite{Dokmanic:2015dr}. Namely, by including only the direct
source, we recover the DS \cite{Tashev:2009vj} and MVDR \cite{Capon:1969da}
beamformers. Options are available to add cancellation of one or more
interferers. Both far and near field formulations can be used. In addition, the
blind source separation algorithm TRINICON \cite{Buchner:2004di} is included.

\subsection{DOA Finding} 
\seclabel{algo_doa}

A base \texttt{DOA} class defines the API of direction finding methods.  The
constructor is responsible for setting the different options of the algorithms.  A
\texttt{locate\_sources} method taking at least one frequency domain frame of the input
signal as argument is responsible for computing the sources locations.
The \texttt{DOA} class is extended to implement several popular algorithms:
the popular multiple signal classification (MUSIC) \cite{Schmidt:1986js} and
steered response power phase transform (SRP-PHAT) \cite{DiBiase:2000uv}, as
well as coherent signal subspace method (CSSM) \cite{Wang:1985jv}, weighted
average of signal subspaces (WAVES) \cite{diClaudio:2001bb}, and test of
orthogonality of projected subspaces (TOPS) \cite{yoon2006tops}.
All implementations cover both localization in 2D and 3D.

\subsection{Adaptive Filtering}
\seclabel{algo_af}

Adaptive filters also share a common structure whereas a base class
\texttt{AdaptiveFilter} defines a simple interface. The constructor is
responsible for passing options of specific algorithms.  A method \texttt{update}
taking a new input sample and a new reference sample updates the current
filter estimate.  The base class is extended to provide implementations of the least
mean squares (LMS), normalized LMS (NLMS), and recursive least squares (RLS) \cite{Haykin:2014tv}.

\subsection{STFT Engine and Real-Time Processing}
\seclabel{algo_rt}

While the \texttt{Beamformer} class includes STFT processing, its
implementation is a one shot, that is it processes the whole signal at once.
This is not suitable for streaming or real-time data sources. A second
implementation of STFT processing is thus given to cover this use case. It is
implemented as an \texttt{STFT} class with the constructor providing the FFT
size, length of zero-padding, windows and other parameters. Methods
\texttt{analysis} and \texttt{synthesis} decompose one frame of the signal into
time-frequency representation and put it back together using overlap-add,
respectively. In between the two, some frequency domain processing is possible.
Options to use efficient FFT libraries such FFTW \cite{Frigo:2005cp} (through
\textit{pyfftw}) or the Intel Math Kernel Library are available.

\section{Conclusion}

We presented the \texttt{pyroomacoustics} Python package for audio processing.
Under an intuitive API, the package includes a small room acoustics simulator
based on the ISM and a number of reference implementations for popular
algorithms for beamforming, DOA finding, and adaptive filtering. A full STFT
engine makes it easy to get started on frame based processing. This
comprehensive set of tools makes it a great starting point for rapidly
prototyping and evaluating new audio processing algorithms.

We plan to continue extending this package in the hope that it can benefit the
audio signal processing community. The current version of
\textit{pyroomacoustics} only supports omnidirectional sources and microphones.
The ability to add directivity patterns to loudspeakers and microphones is
critical to bridge the gap between simulation and experiments. Ideally, both
parametric patterns (e.g.~cardioid microphones) and measured ones should be
supported.

Currently the definition of intricate room geometries is awkward,
especially for non-convex 3D rooms. One way of simplifying this is to implement set
operations of polygons and polyhedra, e.g.~union, difference, etc, making it
possible to build complex shapes from a set of basic ones such as rectangles
and triangles. Another way is to write a parser for files produced by
conventional CAD software (e.g.~SketchUp, AutoCAD).

% -------------------------------------------------------------------------
% Either list references using the bibliography style file IEEEtran.bst
\bibliographystyle{IEEEtran}
\bibliography{bibliography}

\end{sloppy}
\end{document}